\documentclass[12pt]{iopart}
\usepackage{graphicx}
\usepackage{amssymb,stmaryrd,tabularx,setstack}
\usepackage{iopams}

\usepackage{wrapfig}
\usepackage{bm}
\usepackage[T1]{fontenc}
\usepackage[center]{subfigure}
\usepackage[toc,page]{appendix}
\usepackage{float}

\newcommand{\fet}[1]{\boldsymbol{#1}}

\begin{document}
\title{Classical analogies for the force acting on an impurity in a Bose-Einstein condensate}
\author{Jonas R\o nning$^1$, Audun Skaugen$^2$, Emilio Hern\'andez-Garc\'ia$^3$, Crist\'obal L\'opez$^3$, Luiza Angheluta$^1$}

\address{
 $^1$PoreLab, The Njord Centre, Department of Physics, University of Oslo, P. O. Box 1048, 0316 Oslo, Norway}
\address{ 
 $^2$Computational Physics Laboratory, Tampere University, P.O. Box 692, FI-33014 Tampere, Finland}
 \address{
 $^3$IFISC (CSIC-UIB), Instituto de Fisica Interdisciplinar y Sistemas Complejos, 07122 Palma de Mallorca, Spain
 }

\begin{abstract}

We study the hydrodynamic forces acting on a small impurity moving in a two-dimensional Bose-Einstein condensate at non-zero temperature. The condensate is modelled by the damped-Gross Pitaevskii (dGPE) equation and the impurity by a Gaussian repulsive potential coupled to the condensate. For weak coupling, we obtain analytical expressions for the forces acting on the impurity, and compare them with those computed through direct numerical simulations of the dGPE and with the corresponding expressions for classical forces. For non-steady flows, there is a time-dependent force dominated by inertial effects and which has a correspondence in the Maxey-Riley theory for particles in classical fluids. In the steady-state regime, the force is dominated by a self-induced drag. Unlike at zero temperature, where the drag force vanishes below a critical velocity, at low temperatures the impurity experiences a net drag even at small velocities, as a consequence of the energy dissipation through interactions of the condensate with the thermal cloud. This dissipative force due to thermal drag is similar to the classical Stokes' drag. There is still a critical velocity above which steady-state drag is dominated by acoustic excitations and behaves non-monotonically with impurity's speed.

\end{abstract}

\date{\today}

\maketitle

\section{Introduction}
The motion of an impurity suspended in a quantum fluid depends
on several key factors such as the superfluid nature and flow
regime, as well as the size of the impurity and its interaction
with the surrounding fluid~\cite{winiecki2000motion,
wouters2010superfluidity,astrakharchik2004motion,
shukla2016sticking, pinsker2017gaussian}.  Therefore, it is
disputable whether the forces acting on an impurity in a
quantum fluid should bear any resemblance to  classical
hydrodynamic forces. In the case of an impurity immersed in
superfluid liquid helium, classical equations of motion and
hydrodynamic forces are assumed a
priori~\cite{poole2005motion}, since impurities are typically
much larger than the coherence length and then quantum
hydrodynamic effects like the quantum pressure can be
neglected. For Bose-Einstein condensates (BEC) in
dilute atomic gases, impurities can be neutral atoms \cite{chikkatur2000suppression},  ion impurities \cite{zipkes2010trapped,balewski2013coupling} or quasiparticles \cite{jorgensen2016observation}. The size of an impurity in a BEC is typically of the same order of magnitude or smaller than the coherence length,
and quantum hydrodynamic effects cannot be readily ignored.

There are several theoretical and computational studies of the
interaction force between an impurity and a BEC at zero
absolute temperature, using different approaches depending on
the nature of the particle and its interaction with the
condensate. A microscopic approach is used to analyse the
interaction of a rigid particle with a BEC by solving the
Gross-Pitaevskii equation (GPE) for the condensate macroscopic
wavefunction and using boundary conditions such that the
condensate density vanishes at the particle
boundary~\cite{pham2005boundary}. This methodology allows to
study complex phenomena such as vortex nucleation and flow
instabilities, but it is more oriented to find the effects of
an obstacle on the flow rather than the coupled particle-flow
dynamics. In addition, the boundary condition introduces severe
nonlinearities which can only be addressed numerically. At a
more fundamental level of description, the impurity is treated
as a quantum particle with its own wavefunction described by
the Schr\"{o}dinger equation and that is coupled with the GPE
for the macroscopic wavefunction of the
BEC~\cite{berloff2000capture}.  A more versatile model for the
interaction of impurities with the BEC has been explored in
several
papers~\cite{astrakharchik2004motion,shukla2016sticking,griffin2017vortex,shukla2018particles,pinsker2017gaussian,giuriato2019interaction}.
Here, an additional repulsive interaction (a Gaussian or
delta-function potential) is added to model scattering of the
condensate particles with the impurity. The force
on the impurity is determined by this repulsion potential and
the superfluid density through the Ehrenfest theorem. The
strong-coupling limit of this repulsive potential would be
equivalent to the rigid boundary-condition approach. Within
this modelling approach, some works have studied the complex
motion of particles interacting with vortices in the flow, and
the indirect interactions between them arising from the
presence of the fluid
\cite{shukla2016sticking,shukla2018particles}. Another line of
research using this type of modelling focused mainly on the
superfluidity criterion of an equilibrium BEC  ~\cite{astrakharchik2004motion,roberts2006force,roberts2005casimir, pomeau2008hydrodynamic, sykes2009drag,pinsker2017gaussian} and non-equilibrium BEC at zero temperature \cite{pinsker2017beyond}. Within the Bogoliubov perturbation analysis for a small
impurity with weak coupling, analytical expressions can be
derived for the steady-state force on the 
impurity. At zero temperature, this force vanishes below a critical
velocity and corresponds to the dissipationless motion. Above this velocity identified through Landau criterion as the speed of the long-wavelength sound waves, there is a net
drag force and the motion of the impurity is damped by acoustic
excitations. While this is a form of drag, in that the force
opposes motion by dissipating energy, it is not the same as the
classical Stokes' drag in viscous fluids. Recent experiments
probing superfluidity in a BEC are able to indirectly estimate
the drag force by measuring the local heating rate in the
vicinity of the moving laser beam and show that there is
still a critical velocity even at non-zero temperatures and that
the critical velocity is lower for a repulsive potential than
for an attractive one~\cite{singh2016probing}.

In this paper, we study the forces exerted on an impurity
moving in a two-dimensional BEC at low temperature, using an
approach similar to
\cite{astrakharchik2004motion,shukla2016sticking,griffin2017vortex,shukla2018particles,pinsker2017gaussian},
in which a repulsive Gaussian potential is used to describe the
interaction of the particle with the BEC, but using a
dissipative version of the GPE.  Our aim is
to bridge this microscopic approach with the phenomenological
descriptions \cite{poole2005motion} that assume that the forces
from the superfluid are the same as those from a classical
fluid in the inviscid and irrotational case. As in the
classical-fluid case, we find that the force is made of two
contributions: One of them, dominant for very weak
fluid-particle interaction, bears a rather complete analogy
with the corresponding force in classical fluids (inertial or
pressure-gradient force), which depends on local fluid
acceleration and includes the so-called Fax\'{e}n corrections
arising from velocity inhomogeneities close to the particle
position \cite{maxey1983equation}. The difference is that, in a
classical fluid, these corrections arise from the finite size
of the particle and vanish when the particle size becomes zero.
In the BEC, Fax\'{e}n-type corrections arise both from the
particle size (modeled by the range of the particle repulsion
potential) and from the BEC coherence length. As fluid-particle
interaction becomes more important, a second contribution to
the force becomes noticeable, which takes into account the drag
on the particle arising from the perturbation of the flow
produced by the presence of the particle. This is also called the 
particle \emph{self-induced} force. We are able to obtain
explicit formulae for the steady-state 
motion of the particle in an otherwise homogeneous and steady
BEC. This drag is a dissipative (damping) force due to
thermal drag of the BEC with the thermal cloud.
It occurs in addition to the drag due to acoustic excitations
in the condensate that occurs
only above a critical velocity. It can
be compared with the corresponding force in classical fluids,
namely the viscous Stokes drag. We find an analytical expression for this self-induced drag at arbitrary speeds and show that in the low speed limit, it reduces to a linear dependence on speed akin to the classical Stokes drag. 

The rest of the paper is structured as follows. In Sect.
\ref{sec:model}, we discuss the general modelling setup and in
Sect. \ref{sec:perturbation} a perturbation analysis is used to
derive the linearised equations for the perturbations in the
wavefunction related to non-steady condensate flow and the
particle repulsive potential. Subsections \ref{sec:inertial}
and \ref{sec:drag} derive analytical expressions within
perturbation theory for the two contributions to the force
experienced by the particle. In Section \ref{sec:numerics}, we
compare our theoretical predictions with numerical simulations
of the dissipative GPE coupled to the impurity, and the final
section summarises our conclusions.

\section{Modelling approach}
\label{sec:model}

We model the interaction between the impurity and a
two-dimensional (2D) BEC through a Gaussian repulsive potential
which can be reduced to a delta-function limit similar to
previous
studies~\cite{astrakharchik2004motion,pinsker2017gaussian}. The
BEC at low temperatures is well-described by the damped Gross Pitaevskii equation
(dGPE) for the condensate wavefunction $\psi(\fet{r},t)$~\cite{gardiner2002stochastic,rooney2012stochastic,penckwitt2002nucleation,Bradley_2012,Reeves_2013, Reeves_2014,skaugen2016vortex}:
\begin{eqnarray}
\label{eq:GPe}
 i\hbar\partial_t\psi =  (1-i\gamma)\left(-\frac{\hbar^2}{2m}\nabla^2+g|\psi|^2-\mu+ V_{ext}+g_p{\cal U}_p\right)\psi,
\end{eqnarray}
where $g$ is an effective scattering parameter between
condensate atoms. $V_{ext}$ is any external potential used to
confine or stir the condensate. The damping coefficient $\gamma>0$ also called the thermal drag is related to the net exchange of atoms through collisions between thermal atoms with each other or with the condensate at fixed chemical potential $\mu$. In the low-temperature limit, this damping $\gamma$ is very small and can be expressed as a function of temperature $T$~\cite{bradley2008bose,penckwitt2002nucleation}. The dGPE is a phenomenological model that can also be derived from the stochastic GPE in the low temperature limit where noise is negligible \cite{gardiner2002stochastic,rooney2012stochastic}. The dGPE has been used extensively to study different vortex regimes from vortex lattices~\cite{penckwitt2002nucleation} to quantum turbulence ~\cite{Reeves_2014,billam2015spectral,Reeves_2013,Bradley_2012,neely2013characteristics}  and was shown to capture well, at least qualitatively, experimental observations ~\cite{neely2013characteristics}. 

 A hydrodynamic description of the BEC uses the Madelung transformation of the
wavefunction $\psi=|\psi|e^{i\phi}$ to define the condensate 
density as $\rho(\fet r,t)=|\psi(\fet r,t)|^2$ and the
condensate velocity as $\fet v(\fet r,t)=(\hbar/m) \nabla
\phi(\fet r,t)$. This velocity can also be obtained from the
superfluid current $\fet J(\fet r,t)$ as
\begin{equation}
\fet J = \frac{\hbar}{2 m i} \left(\psi^*\nabla\psi - \psi\nabla\psi^* \right) =\rho \fet v,  
\label{eq:current}
\end{equation}
where $\psi^*$ denotes the complex conjugate of $\psi$. In addition
to damping the BEC velocity, the presence of $\gamma\neq 0$ in the
dGPE also singles out the value $\rho_h=|\psi|^2=g/\mu$ as
the steady homogeneous density value when the phase is constant
and $V_{ext}=0$.

The interaction potential ${\cal U}_p(\fet r-\fet{r}_p)$
between the condensate and the impurity is modelled by a
Gaussian potential ${\cal U}_p(\fet r-\fet{r}_p)=
\mu/(2\pi \sigma^2) e^{-(\fet r-\fet
r_p)^2/(2\sigma^2)}$.  The parameter $g_p>0$ is the weak coupling constant
for repulsive impurity-condensate interaction, $\fet r_p=\fet
r_p(t)$ denotes the center-of-mass position of the impurity,
and $\sigma$ its effective size. Here we consider an impurity
of size $\sigma$ of the order the coherence length
$\xi=\hbar/\sqrt{m\mu}$ of the condensate. The impurity is too
small to nucleate vortices in its
wake~\cite{reeves2015identifying}. Instead, it will
create acoustic excitations with Bogoliubov-\u{C}erenkov wake. 
Similar wave fringes in the 
condensate density have been reported numerically
in~\cite{wouters2010superfluidity} for a different realisation
of non-equilibrium Bose-Einstein condensates. In the limit of a
point-like impurity, the Gaussian interaction potential
converges to a two-body scattering potential  ${\cal U}_p(\fet
r-\fet{r}_p,t)= \mu \delta(\fet r-\fet r_p(t))$ that has been
used in previous analytical
studies~\cite{astrakharchik2004motion,shukla2016sticking,griffin2017vortex,shukla2018particles}.
Note that we are modelling only the interaction of the particle
with the BEC, so that the viscous-like drag we obtain
arises from the indirect coupling to the thermal bath via the
BEC. Any direct interaction of the particle with the thermal
cloud will lead to additional forces which could be important at high temperatures, and which are not included here. 

In order to gain insight into the forces and their
relationship with the classical case, we keep the set-up as
simple as possible. We consider a 2D 
condensate and assume that the size of the condensate is large enough so that
we can neglect inhomogeneities in the confining part of
$V_{ext}$.  Also, we consider a
neutrally buoyant impurity so that effects of gravity can be
neglected. This would imply $V_{ext}=0$ except if an external
forcing is introduced to stir the system, in which case we
assume the support of this external force is sufficiently far
from the impurity.

The impurity and the condensate will exert an interaction force
on each other that is determined by the Ehrenfest theorem for
the evolution of the center-of-mass momentum of the particle.
The potential force $-g_p\nabla {\cal U}_p(\fet r-\fet{r}_p)$
is the force exerted by an impurity on a condensate particle at
position $\fet r$.  By space averaging over condensate density,
we then determine the force exerted by the impurity on the
condensate as $-g_p\int d\fet r |\psi(\fet r,t)|^2 \nabla {\cal
U}_p(\fet r-\fet{r}_p)$ ~\cite{shukla2018particles}. Hence, the
force acting on the impurity has the opposite sign and is equal
to
\begin{eqnarray}\label{eq:fp0}
\fet F_p(t) &=& + g_p\int d^2\fet r |\psi(\fet r,t)|^2 \nabla {\cal U}_p(\fet r-\fet{r}_p)
\end{eqnarray}
which, through an integration by parts, is equivalent to
\begin{eqnarray}\label{eq:fp1}
\fet F_p(t) &=& -g_p\int d^2\fet r  ~{\cal U}_p(\fet r-\fet{r}_p ,t)  \nabla |\psi(\fet r,t)|^2 .
\label{eq:forcegradrho}
\end{eqnarray}
Note that this last expression can also be used, reversing the
sign, to give the force exserted on the BEC by a laser of beam
profile given by ${\cal U}_p$.

At zero temperature, i.e. $\gamma=0$, and if we neglect the effect
of quantum 
fluctuations~\cite{roberts2005casimir,roberts2006force,pomeau2008hydrodynamic,sykes2009drag},
the impurity moves without any drag through a uniform
condensate below a critical velocity, which is the
low-wavelength speed of sound $c=\sqrt{\mu/m}$, as determined
by the condensate linear excitation spectrum, in agreement with
Landau's criterion of
superfluidity~\cite{astrakharchik2004motion}. Above the
critical speed, the impurity will create excitations, and
depending on the size of the impurity these excitations range
from acoustic waves (Bogoliubov excitation spectrum) to vortex
dipoles and to von-Karman street of vortex
pairs~\cite{reeves2015identifying}. Previous studies focused on
the theoretical investigations of the self-induced drag force
and energy dissipation rate in the presence of Bogoliubov
excitations emitted by a pointwise
~\cite{astrakharchik2004motion,roberts2006force,sykes2009drag}
or finite-size~\cite{pinsker2017gaussian} particle, or
numerical investigations of the drag force due to vortex
emissions~\cite{winiecki2000motion,griffin2017vortex,shukla2018particles}.
The energy dissipation rate depends on whether the impurity is
heavier, neutral or lighter with respect to the mass of the
condensate particles~\cite{shukla2018particles}. The dependence
on the velocity of the self-induced drag force above the
critical velocity changes with the spatial
dimensions~\cite{astrakharchik2004motion}. This means that the
energy dissipation rate is also dependent on the spatial
dimensions.  If instead of a single impurity one considers many
of them there will be, besides direct inter-particle
interactions, additional forces between the impurities mediated
by the flow, leading to a much more complex many-body dynamics
even in an otherwise uniform condensate, as discussed in
\cite{shukla2016sticking}. Here we neglect all these effects
and consider a single impurity in a 2D BEC.

We rewrite the dGPE in dimensionless units by using the
characteristic units of space and time in terms of the
long-wavelength speed of sound $c=\sqrt{\mu/m}$ in the
homogeneous condensate and the coherence length $\xi=\hbar/(m
c)=\hbar/\sqrt{m\mu}$. Space is rescaled as $\fet r\rightarrow
\tilde{\fet r} \xi$ and time as $t\rightarrow \tilde t \xi/c$.
In addition, the wavefunction is also rescaled $\psi
\rightarrow \tilde \psi\sqrt{\mu/g}$, where $g/\mu$ is the
equilibrium particle-number density corresponding to the
solution with constant phase if $V_{ext},\mathcal U_p=0$. The
external potential, $V_{ext}=\mu \tilde V_{ext}$, and the
interaction potential, $g_p \mathcal U_p = \mu \tilde g_p
\tilde{\mathcal U_p}$, are measured in units of the chemical
potential $\mu$ with $\tilde{\mathcal U_p} = 1/(2\pi a^2)
e^{-({\fet \tilde r}-{\fet \tilde r}_p)^2/(2a^2)}$, and $a =
\sigma/\xi$, $\tilde{g}_p = g_p/(\xi^2\mu)$. 
Henceforth, the dimensionless form of the dGPE reads as
\begin{eqnarray}
\label{eq:GPe_dimless}
\tilde \partial_t \tilde\psi =
(i+\gamma)\left(\frac{1}{2}\tilde\nabla^2+1-\tilde V_{ext}-\tilde g_p\tilde{\mathcal U_p}-|\tilde\psi|^2\right)\tilde\psi  .
\end{eqnarray}

In these dimensionless units, the force
from Eq. (\ref{eq:forcegradrho}) exerted on an impurity reads as $\fet{F}_p=
(\mu^2\xi/g) \tilde{\fet{F}}_p$, where
\begin{eqnarray}
\tilde{\fet{F}}_p(t) =
-\tilde g_p\int d^2\tilde{\fet r} \tilde{\mathcal U_p}(\fet r-\fet{r}_p)\tilde\nabla |\tilde\psi(\tilde{\fet r},t)|^2. \label{eq:fp2}
\end{eqnarray}
For the rest of the paper, we will now omit the tildes over the
dimensionless quantities.

In the limit of a point-like particle, $\mathcal U_p=
\delta(\fet r-\fet{r}_p)$, the force from Eq. (\ref{eq:fp2})
becomes
\begin{eqnarray}
\fet{F}_p(t) = -g_p\nabla|\psi(\fet r,t)|^2 |_{\fet r=\fet r_p(t)}.
\label{eq:fp3}
\end{eqnarray}
%

\section{Perturbation analysis}
\label{sec:perturbation}

For a weakly-interacting impurity, the condensate wavefunction
$\psi$ can be decomposed into an unperturbed wavefunction
$\psi_0(\fet r)$ describing the motion and density of the fluid
in the absence of the particle and the perturbation
$\delta\psi_1(\fet r)$ due to the impurity's repulsive
interaction with the condensate, hence $\psi = \psi_0+g_p
\delta\psi_1$.  Weak particle-condensate interaction condition is that $\max(g_p \mathcal{U}_p) \ll 1$, or $g_p \ll 2\pi a^2$, which means that
the particle-condensate interaction of strength $g_p$ and
range $\sigma$ is small compared with the energy scale
given by the chemical potential $\mu=1$ (dimensionless units). 

The unperturbed wavefunction $\psi_0(\fet r,t)$
can be spatially-dependent, if it is initialised in a
nonequilibrium configuration, or if external forces
characterised by $V_{ext}$ are at play. Here, we consider
deviations with respect to the steady and uniform equilibrium
state ($\psi_h=1$ in dimensionless units). As
stated before, we do not consider large extended
inhomogeneities produced by a trapping potential, and assume
that any stirring force acting on the BEC is far from the
particle. Thus, we treat inhomogeneities close to the particle
as small perturbations to the uniform state $\psi_h=1$:
 $\psi_0(\fet r,t)=1+\delta\psi_0(\fet
r,t)$. Combining the two types of perturbations, and using the
relationships of the wavefunction to the density, velocity and
current (Eq. (\ref{eq:current}), which in dimensionless units
reads $\rho\fet v=(\psi^*\nabla\psi-\psi\nabla\psi^*)/(2i)$) we
find
\begin{eqnarray}
\label{eq:density_perturb}
\psi &=& 1 +\delta \psi_0 +g_p \delta\psi_1   \\
\rho &=& 1 +\delta \rho_0+g_p \delta\rho_1,   \\
\fet v &=& \delta\fet v^{(0)} + g_p \delta \fet v^{(1)},
\end{eqnarray}
where
\begin{eqnarray}
&&\delta \rho_0 =
\delta\psi_0+\delta\psi_0^*, \quad  \delta\rho_1 =
\delta\psi_1+\delta\psi_1^* ,\label{eq:density_linear} \\
&&\delta \fet v^{(0)} = \frac{1}{2 i}\nabla \left(\delta \psi_0 - \delta \psi_0^*\right), \quad
\delta\fet v^{(1)} =\frac{1}{2 i}\nabla (\delta\psi_1 - \delta\psi_1^*).\nonumber\\
\label{eq:velocity_linear}
\end{eqnarray}

Combining Eq. (\ref{eq:fp2}) with the expressions for the
density perturbations, we have that the total force can be
split into the contribution from the density variations in the
BEC by causes external to the particle (initial preparation,
stirring forces in $V_{ext}$, ...), and the density
perturbations due to the presence of the particle $\fet{F}_p=
\fet F^{(0)}+\fet F^{(1)}$:
\begin{eqnarray}
\fet{F}^{(0)}(t) = - \frac{g_p}{2\pi a^2} \int d^2\fet r e^{- \frac{(\fet r-\fet r_p(t))^2}{2a^2}}
\nabla \delta\rho_0(\fet r,t),
\label{eq:fp_unperturb}
\\
\fet{F}^{(1)}(t) = - \frac{g_p^2}{2\pi a^2} \int d^2\fet r e^{- \frac{(\fet r-\fet r_p(t))^2}{2a^2}}
\nabla \delta\rho_1(\fet r,t).
\label{eq:fp_perturb}
\end{eqnarray}
The perturbative splitting of the force in these two
contributions is completely analogous to the corresponding
classical-fluid case in the incompressible
\cite{maxey1983equation} and in the compressible
\cite{parmar2012equation} situations. The $\fet F^{(0)}$
contribution is the equivalent to the classical inertial or
pressure-gradient force on a test particle, which does not
disturb the fluid, in a inhomogeneous and unsteady flow. We call this the \emph{inertial} force. The
$\fet F^{(1)}$ contribution takes into account perturbatively
the modifications on the flow induced by the presence of the
particle, and it is called the \emph{self-induced drag} on
the particle. To complete the comparison with the classical
expressions \cite{maxey1983equation,parmar2012equation}, we
need to express Eqs. (\ref{eq:fp_unperturb}) and
(\ref{eq:fp_perturb}) in terms of the unperturbed velocity
field $\fet v^{(0)}(\fet r,t)=\delta\fet v^{(0)}(\fet r,t)$ and
of the particle speed $\fet V_p(t)=\fet{\dot r}_p(t)$. We are
able to do so in a general situation for the inertial force
$\fet F^{(0)}$. For $\fet F^{(1)}$, we obtain analytical
expressions in the simple case where the impurity is moving
with a constant velocity in an otherwise uniform BEC.

The desired relationships between $\nabla \delta\rho_0$ and
$\nabla \delta\rho_1$ in Eqs.
(\ref{eq:fp_unperturb})-(\ref{eq:fp_perturb}), and $\delta\fet
v^{(0)}$ and $\fet V_p$ will be obtained from the linearization
of the dGPE Eq. (\ref{eq:GPe_dimless}) around the uniform
steady state $\psi_h=1$:

\begin{eqnarray}
\partial_t \delta\psi_0  &=& (i+\gamma)\left(\frac{1}{2}\nabla^2 -1\right)\delta\psi_0 
- (i+\gamma)\delta\psi^*_0,
\label{eq:psi0lin} \\
\partial_t \delta\psi_1  &=&
(i+\gamma)\left(\frac{1}{2}\nabla^2-1\right)\delta\psi_1
-(i+\gamma)\delta\psi_1^* - (i+\gamma) \mathcal U_p(\fet r-\fet r_p) \ .
\label{eq:psi1lin}
\end{eqnarray}

Terms containing $V_{ext}$ are not included in Eq.
(\ref{eq:psi0lin}) because of our assumption of sufficient
distance between possible stirring sources and the neighborhood
of the particle position, the only region that--as we will
see-- will enter into the calculation of the forces. In the
next sections we solve these linearised equations to relate
density perturbations to undisturbed velocity field and
particle velocity.

\subsection{Inertial force} \label{sec:inertial}

To convert Eq. (\ref{eq:fp_unperturb}) for the inertial force
into an expression suitable for comparison with the
corresponding term in classical fluids, we need to express
$\nabla\delta\rho_0$ in terms of the undisturbed velocity field
$\fet v^{(0)}(\fet r,t)=\delta\fet v^{(0)}(\fet r,t)$. To this
end, we substract Eq. (\ref{eq:psi0lin}) from its complex
conjugate, obtaining:
\begin{eqnarray}
\left(\nabla^2-4\right)\nabla\delta\rho_0=
4 \left(\partial_t-\frac{\gamma}{2}\nabla^2\right)\delta\fet v^{(0)},
\label{eq:drhodv0}
\end{eqnarray}
where we have used Eqs. (\ref{eq:density_linear}) and
(\ref{eq:velocity_linear}). Since the force formulae require to
obtain the condensate density in a neighbourhood of the particle
position, it is convenient to move to frame co-moving with the particle. Thus we change variables from $(\fet r, t)$ to
$(\fet z, t)$, with $\fet z=\fet r -\fet r_p(t)$, and the
velocity field will be now referred to the particle velocity
$\fet V_p(t)=\fet{\dot r}_p(t)$: $\delta \fet w^{(0)}(\fet
z,t)=\delta\fet v^{(0)}(\fet r,t)- \fet V_p(t)$. Equation
(\ref{eq:drhodv0}) becomes:
\begin{eqnarray}
\left(\nabla_z^2 - 4\right)\nabla_z\delta\rho_0 = 
4 \left(\partial_t-\fet V_p\cdot\nabla_z-\frac{\gamma}{2}\nabla_z^2\right)\delta\fet w^{(0)} + \fet{\dot V}_p(t),
\label{eq:drhodv}
\end{eqnarray}
which has the corresponding equation for its Green's function given by
\begin{equation}
\left(\nabla_z^2 - 4\right)G(\fet z) = \delta(\fet z)
\end{equation}
with the boundary condition $G(|\fet
z|\rightarrow\infty)\rightarrow 0$ (corresponding to vanishing
$\nabla_z \delta\rho_0(\fet r)$ at $|\fet r|=\infty$). The
solution is given by the zeroth order modified Bessel function
$G(\fet z)=-K_0(2|\fet z|)/(2\pi)$. Hence, the gradient of the
density perturbation can be written as the convolution with the
Green's function:
\begin{equation}
\fl\nabla_z\delta\rho_0(\fet z,t)=
-\frac{2}{\pi}\int d\fet z' K_0(2|\fet z-\fet z'|)
\left[\left(\partial_t-\fet V_p\cdot\nabla_{\fet z'}-\frac{\gamma}{2}\nabla^2_{\fet z'}\right)\delta\fet w^{(0)}(\fet z',t) + \fet{\dot V}_p(t) \right] ,
\label{eq:nablarho}
\end{equation}
and the expression for the force (\ref{eq:fp_unperturb}), using
the comoving variables $(\fet z,t)$, becomes:
\begin{equation}
\fl\fet F^{(0)}(t)=
-\frac{g_p}{\pi^2a^2}\int d\fet z e^{-\frac{z^2}{2 a^2}} \int d\fet z' K_0(2|\fet z-\fet z'|)
\left[\left(\partial_t-\fet V_p\cdot\nabla_{\fet z'}-\frac{\gamma}{2}\nabla^2_{\fet z'}\right)\delta\fet w^{(0)}(\fet z',t) + \fet{\dot V}_p(t) \right] .
\label{eq:F0full}
\end{equation}

The above expression is a weighted average of contributions
from properties of the fluid velocity in a neighborhood of the
impurity center-of-mass position ($\fet z = 0$ in the comoving
frame). The size of this neighborhood is given by the
combination of the range of the Bessel function kernel, which
in dimensional units would be the correlation length $\xi$, and
the range of the Gaussian potential, $a$, giving an effective
particle size. In classical fluids, the analogous force on a
spherical particle involves the average of properties of the
undisturbed velocity field within the sphere size
\cite{parmar2012equation}, and there is no equivalent to the
role of $\xi$.

As in the classical
case~\cite{maxey1983equation,parmar2012equation}, if fluid
velocity variations are weak at scales below $a$ and $\xi$, we
can approximate the condensate velocity by a Taylor expansion
near the impurity, i.e.:
\begin{equation}
\delta w_i^{(0)}(\mathbf z',t)\approx \delta
w_i^{(0)}(t)+\sum_j
e_{ij}(t) z'_j 
+\frac{1}{2}\sum_{jk}e_{ijk}(t)  z'_j z'_k + \ldots,
\end{equation}
where the indices $i,j,k=x,y$ denote the coordinate components.
$e_{ij}(t)=\partial_j \delta w_i^{(0)}(\mathbf z,t)|_{\fet
z=0}$ and $e_{ijk}(t)=\partial_j
\partial_k \delta w_i^{(0)}(\mathbf z,t)|_{\fet z=0}$ are
gradients of the unperturbed condensate relative velocity.
Inserting this expansion into Eq. (\ref{eq:F0full}), and
performing the integrals of the Gaussian and of the Bessel
function (using for example $\int K_0(2|\fet z|)d\fet z=\pi/2$
and $\int z_i z_j K_0(2|\fet z|)d\fet z
=(\delta_{ij}/2)\int_0^\infty 2\pi z^3
K_0(2z)dz=\delta_{ij}\pi/4$), we obtain:
\begin{equation}
\fl\fet F^{(0)}(t) \approx g_p \fet{\dot V}_p(t)+ g_p\left[\partial_t
    - \fet{V}_p(t)\cdot\nabla_z +\frac{a^2}{2}\partial_t\nabla^2_z  
- \frac{\gamma}{2}\nabla_z^2 +
    \frac{1}{4}\partial_t\nabla^2_z  \right] \delta\fet w^{(0)}(\fet z,t)|_{\fet z=\fet 0} \ .
    \label{eq:Fi_th}
\end{equation}

The terms containing Laplacians are analogous to the Fax\'{e}n
corrections in classical fluids~\cite{maxey1983equation} which
arise for particles with finite size. Here, they arise from a
combination of the finite effective size of the particle, $a$,
and of the quantum coherence length, $\xi=1$. This last effect
remains even in the limit of vanishing particle size
$a\rightarrow 0$. Interestingly, one of the two terms in these
quantum corrections depend on $\gamma$ hence indirectly on the
presence of the thermal cloud.

As in the classical case, if flow inhomogeneities are
unimportant below the scales $a$ and $\xi$, we can neglect the
Laplacian terms in Eq. (\ref{eq:Fi_th}). Returning to the
variables $(\fet r,t)$ in the lab frame of reference, the terms
containing $\fet V_p$ cancel out, showing that the inertial
force is mainly given by the local fluid acceleration:
\begin{eqnarray}\label{eq:pd1}
\fet F^{(0)}(t) = g_p \partial_t \delta\fet v^{(0)}(\fet r,t) \big|_{\fet r=\fet r_p(t)}.
 \end{eqnarray}
We have assumed a small non-uniform unperturbed velocity field
$\fet v^{(0)}(\fet r,t)= \delta\fet v^{(0)}(\fet r,t)$. To
leading order in velocity, the partial derivative
$\partial_t\delta\fet v^{(0)}$ and the material derivative
$D\delta\fet v^{(0)}/Dt=\partial_t\delta\fet v^{(0)}+\delta\fet
v^{(0)}\cdot \nabla \delta\fet v^{(0)}$ are identical. In
classical fluids the same ambiguity occurs and it has been
established, on physical grounds and by going beyond
linearisation, that using the material derivative is more
correct \cite{maxey1983equation}. After all, using this
material derivative in the equation of motion simply means
that, under the above approximations and in places where
stirring and other external forces are absent, the local
acceleration on the impurity arises from the corresponding
acceleration of the condensate. Since for $a\rightarrow 0$ the
condensate-impurity interaction has a similar scattering
potential (delta function) as that for the interaction between
condensate particles, similar accelerations would be
experienced by a condensate particle and by the impurity, just
modulated by a different coupling constant. Thus, replacing
$\partial_t$ by $D/Dt$ in (\ref{eq:pd1}) the approximate
inertial force becomes:
\begin{eqnarray}\label{eq:pd}
\fet F^{(0)}(t)  &=&
g_p \left.\frac{D \fet v^{(0)}}{Dt}\right|_{\fet r=\fet r_p(t)}\ ,
\label{eq:InertialSimpleDimless}
 \end{eqnarray}
or, if we return back to dimensional variables:
\begin{eqnarray}\label{eq:pd_dim}
\fet F^{(0)}(t)  &=&
\frac{g_p}{g}  m \left.\frac{D \fet v^{(0)}}{Dt}\right|_{\fet r=\fet r_p(t)}\ .
\label{eq:InertialSimpleDim}
 \end{eqnarray}
This is equivalent to the equation for the inertial force in
classical fluids~\cite{maxey1983equation} except that the
coefficient of the material derivative in the classical case is
the mass of the fluid fitting in the size of the impurity. In
the comoving frame, replacement of the partial by the material
derivative amounts to replace $(\partial_t-\fet V_p\cdot
\nabla_{\fet z'})\delta\fet w^{(0)}$ in Eq.
(\ref{eq:F0full}) by $D\delta\fet w^{(0)}/Dt$. Eq.
(\ref{eq:pd}) is expected to be valid for small values of $g_p$
and in regions where fluid velocity and density inhomogeneities
are both small and weakly varying. 
At this level of
approximation neither compressibility nor dissipation effects
appear explicitly in the inertial force, in analogy with
classical compressible fluids~\cite{parmar2012equation}. But
these effects are indirectly present by determining the
structure of the field $\fet v^{(0)}(\fet r,t)$. 

\subsection{Self-induced drag force} \label{sec:drag}

The consideration of the self-induced force on
a particle moving through a classical fluids leads to
different terms, namely
\cite{maxey1983equation,parmar2012equation} the viscous
(Stokes) drag, the unsteady-inviscid term that in the
incompressible case becomes the added-mass force, and the
unsteady-viscous term that in the incompressible case becomes
the Basset history force. They are expressed in terms of the
undisturbed velocity flow $\fet v^{(0)}$ and the particle
velocity $\fet V_p(t)$. Here, for the BEC case, we are able to
obtain the self-induced force only for a particle moving at
constant speed on the condensate. For the classical fluid case,
in this situation the only non-vanishing force is the Stokes
drag, so that this is the force we have to compare our
result with. We note that the condensate itself in the absence
of the particle perturbation can be in any state of (weak)
motion since in our perturbative approach summarised in Eqs
(\ref{eq:psi0lin})-(\ref{eq:psi1lin}), the inhomogeneity
$\delta\psi_0$ and the $g_p$-perturbation $\delta\psi_1$ are
uncoupled.

It is convenient to transform the problem to the co-moving frame $(\fet r,t)\rightarrow (\fet
z,t)$ with $\fet z=\fet r - \fet r_p(t)$, so that Eq.
(\ref{eq:psi1lin}) becomes
\begin{equation}
\partial_t \delta\psi_1 - \fet{V}_p \cdot \nabla \delta\psi_1 =
(i+\gamma)\left(\frac{1}{2}\nabla^2-1\right)\delta\psi_1
-(i+\gamma)\delta\psi_1^* - (i+\gamma) \mathcal U(\fet r) \ .
\label{eq:psi1linComoving}
\end{equation}
Note that such Galilean transformations of the GPE using a
constant $\fet V_p$ are often accompanied by a multiplication
of the transformed wavefunction by a phase factor $\exp(i\fet
V_p \cdot \fet z + \frac i 2 V_p^2 t)$, in order to transform
the condensate velocity (see below) to the new frame of
reference, and account for the shift in kinetic energy. Indeed,
such a combined transformation leaves the GPE unchanged at
$\gamma = 0$~\cite{pismen1999} (but not for $\gamma>0$). The
density perturbation $\delta\rho_1$ is already given correctly
by $\delta\psi_1+\delta\psi_1^*$, where $\delta\psi_1(\fet
z,t)$ is the solution of (\ref{eq:psi1linComoving}), without
the need of any additional phase factor. The velocity in the
co-moving frame would need to be corrected as
$\delta\omega^{(1)}(\fet z,t)=\delta \fet v^{(1)}-\fet V_p$,
with $\delta v^{(1)}$ given by Eq.
(\ref{eq:velocity_linear}) in terms of the solution of
Eq. (\ref{eq:psi1linComoving}).

Eq. (\ref{eq:psi1linComoving}) in the steady-state can be
solved by using the Fourier transform $\delta\psi_1(\fet z)
=1/(2\pi)^2 \int d^2\fet k e^{i\fet k\cdot \fet z}
\delta\hat\psi_1(\fet k)$. It follows that the linear system of
equations for $\delta\hat\psi_1(\fet k)$ and
$\delta\hat\psi_1^*(-\fet k)$ is given by
\begin{eqnarray}
\left[-2i \fet k\cdot\fet V_p +(i+\gamma) (k^2+2)\right]&\delta\hat\psi_1& + 2(i+\gamma)\delta\hat\psi_1^* = 
- 2(i+\gamma) e^{-\frac{a^2k^2}{2}} ,\nonumber\\
\left[-2i \fet k\cdot\fet V_p +(-i+\gamma) (k^2+2)\right]&\delta\hat\psi_1^*& + 2(-i+\gamma)\delta\hat\psi_1 =  
- 2(-i+\gamma) e^{-\frac{a^2k^2}{2}} .\nonumber\\
\end{eqnarray}
By solving these equations, we find $\delta\hat\psi_1(\fet k)$
and $\delta\hat\psi_1^*(-\fet k)$, and the  Fourier transform
of the density perturbation $\delta\rho_1 =
\delta\psi_1^*+\delta\psi_1$ then follows as
\begin{equation}
    \delta\hat\rho_1 = \frac{e^{-\frac{k^2a^2}{2}} (4 k^2(1+\gamma^2)-8i\gamma\fet{k}\cdot \fet{V}_p)}{4\fet{k}\cdot\fet{V}_p(\fet{V}_p \cdot \fet{k}+i\gamma k^2 +2i\gamma)- k^2(4+k^2)(1+\gamma^2)}.
    \label{Rho1Delta}
\end{equation}
Using the convolution theorem, we can express the self-induced
force (\ref{eq:fp_perturb}) (in the co-moving frame, i.e. with
$\mathbf r_p=0$) in terms of $\delta\hat\rho_1$ as
\begin{equation}
    \fet{F}^{(1)}= - \frac{g_p^2}{(2\pi)^2}\int d^2\fet k i\fet{k}\delta\hat{\rho}_1(\fet{k}) e^{-\frac{k^2a^2}{2}} .
\end{equation}
This force can be decomposed into the normal and tangential
components relative to the particle velocity $\fet V_p$: $ \fet
F^{(1)} = F_{\|} \fet e_{\|}+F_{\perp} \fet e_{\perp}$. Due to
symmetry, the normal component vanishes upon polar integration,
and we are left with the tangential, or drag, force
\begin{equation}
\fl F_{\|}= -\frac{g_p^2}{(2\pi)^2} \int_0^\infty dk\int_0^{2\pi} d\theta e^{-k^2a^2}\frac{i k^2\cos(\theta) \left[4 k^2(1+\gamma^2) - 8i\gamma kV_p\cos(\theta)\right]}{4kV_p\cos(\theta)(kV_p\cos(\theta)+ i\gamma k^2 + 2i\gamma) - k^2(4 + k^2)(1+\gamma^2)}.
\label{eq:force_int}
\end{equation}
$V_p$ is the modulus of $\fet V_p$. At zero temperature, i.e.
when $\gamma=0$, the drag force reduces to the one that has
also been calculated for a point particle in
Refs.~\cite{astrakharchik2004motion} and in
\cite{pinsker2017gaussian} for a finite-$a$ particle:
\begin{equation}
    F_\parallel = - \frac{g_p^2}{\pi^2}\int_0^\infty dk\int_0^{2\pi} d\theta\frac{i k^2 \cos{\theta}e^{-k^2a^2}}{4V_p^2 \cos^2{\theta}- (4+k^2)} \ ,
\end{equation}
which is zero for particle speed smaller than the critical
value given by the long-wavelength sound speed, $V_p < c=1$.
Above the critical speed, the integral has poles and acquires a
non-zero value given by
\begin{equation}
    F_\parallel = -\frac{g_p^2 k_{max}^2}{4V_p}  e^{-\frac{a^2 k_{max}^2}{2}} \left[I_0\left(\frac{a^2 k_{max}^2}{2}\right) - I_1\left(\frac{a^2 k_{max}^2}{2}\right) \right]
    \label{eq:forcegamma0}
\end{equation}
in terms of the modified Bessel functions of the first kind
$I_n(x)$ and where $k_{max} =2\sqrt{V_p^2 -1}$. For vanishing
$a$ the dominant term is proportional to $(V_p^2-1)/V_p$
~\cite{astrakharchik2004motion}. This drag is pertaining to
energy dissipation by radiating sound waves in the condensate
away from the impurity. We emphasise again that $a$ is
small enough such that emission of other excitations, such as
vortex pairs, does not occur. It is important to note
\cite{astrakharchik2004motion,pinsker2017gaussian} that in
order to obtain a real value for the force in Eq.
(\ref{eq:forcegamma0}) one has to consider that it has been
obtained from the limit $\gamma\rightarrow 0^+$ in
(\ref{eq:force_int}), which implies that an infinitesimal
positive imaginary part needs to be considered in the
denominator to properly deal with the poles in the integral.

\begin{figure}[t]
\includegraphics[width=\textwidth]{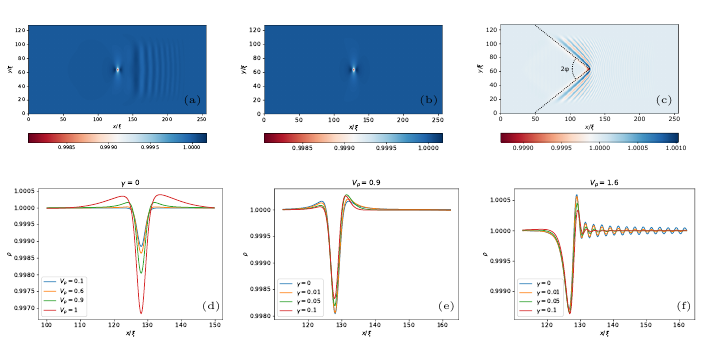}
\caption{Panels (a)-(c) show 2D snapshots of the condensate
density for $\gamma=0$. The impurity is at
$x/\xi=128$ and $y/\xi=64$. (a) is at  $V_p=0.9$ and at time $t=200$, with transient waves still in the
system. (b) is for the same $V_p=0.9$ and at
$t=2000$, when the final steady state has been reached. Panel (c) is for $V_p=1.6>1$,
for which some waves remain attached to the impurity as front
fringes and the Bogoliubov-\u{C}erenkov wake with $\sin(\phi)= 1/V_p$. Panels (d-f) show cross-section
profiles along the $x$ direction of the steady-state condensate
density around the impurity. Panel (d) shows the front-rear
symmetry of the steady profiles when $V_p\leq 1$ and
$\gamma=0$. An asymmetry develops (panel (e)) for $\gamma>0$,
which relates to the net viscous-like drag. Panel (f) displays
density profiles for $V_p=1.6>1$ and different values of
$\gamma$. The asymmetric density profile corresponds to waves
trapped in front of the moving particle. With increasing
$\gamma$, these waves are damped out.} \label{fig:density_gamma}
\end{figure}

In general, for a non-zero $\gamma$, Eq. (\ref{eq:force_int})
simplifies upon an expansion in powers of $V_p$ to the leading
order. For the linear term in $V_p$, we can perform the polar
integration and arrive at
\begin{equation}
F_\parallel = -\frac{2}{\pi}V_p \frac{\gamma}{1+\gamma^2}g^2_p \int \frac{k^3 e^{-a^2k^2}}{(4 +k^2)^2}dk \ .
\end{equation}
Substituting $u = a^2(k^2 + 4)$, we find
\begin{eqnarray}
  F_{\|}  =-V_p \frac{\gamma}{1+\gamma^2}g_p^2 \frac{1}{\pi} \left[ e^{4a^2 }E_1(4a^2)(1+4a^2) - 1\right], \nonumber\\
\label{eq:drag_coeff_full}
\end{eqnarray}
where $E_1(x)$ denotes the positive exponential integral. When $a \to 0$, the expression inside the bracket diverges as
$-\gamma_E - 1 - \ln(4a^2)$
with $\gamma_E$ begin the
Euler-Mascheroni constant. It is therefore necessary to keep a finite size $a$.

This drag force is akin to the viscous Stokes drag 
in classical fluids, but it is due to loss of energy in the condensate through its interaction with the thermal cloud. 
 The effective drag coefficient depends on the
thermal drag such that it vanishes at zero temperature. But it
also depends non-trivially on the size of the impurity and it
diverges in the limit of point-like particle. Fax\'{e}n
corrections involving derivatives of the unperturbed flow are
not present here because of the decoupling between
$\delta\psi_0$ and $\delta\psi_1$ arising in the perturbative
approach leading to (\ref{eq:psi0lin})-(\ref{eq:psi1lin}).

\section{Numerical results}\label{sec:numerics}

\begin{figure}[t]
\centering
\includegraphics[width=\textwidth]{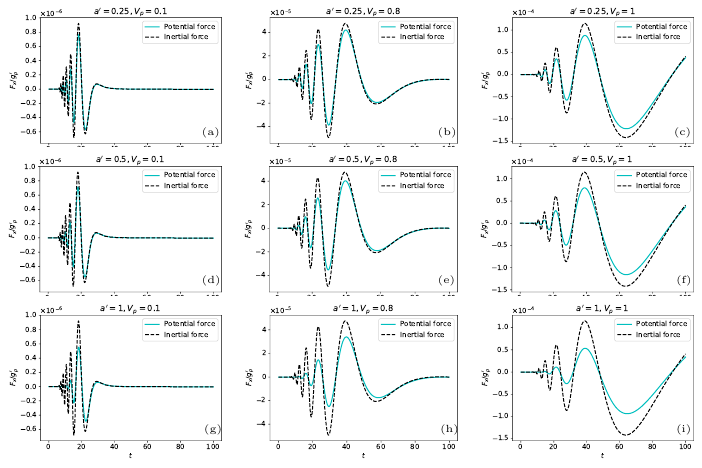}
\caption{$x$ component of the time-dependent force $F_x/g'_p$, using direct numerical
simulations of the dGPE Eq. (\ref{eq:ComovingdGPE}), on a test
particle of size $a'=0.25, 0.5, 1$ at a relative position
$(\Delta x,\Delta y)=(10,20)$ with respect to the position of
the particle producing the flow perturbation. The speed of both
particles is $V_p = 0.1, 0.8, 1$, and $\gamma=0$. Cyan
continuous lines correspond to the full force ($x$ component) from the exact
expression Eq. (\ref{eq:fp2}). They are labeled as `potential
force' because of the rather explicit appearance of the
interaction potential in this formula. Black dotted lines are
the predictions for the inertial force from the approximation
Eq. (\ref{eq:InertialSimpleDimless}) (computed in the comoving
frame as explained in the text). } \label{fig:inertial_force}
\end{figure}

To test the analytical predictions of the inertial force and
the self-induced drag deduced above from the total force
expression Eq. (\ref{eq:fp2}), we performed numerical
simulations of the dGPE.  Actually, our simulations are done in
the co-moving frame of the impurity moving at constant velocity
$\fet V_p$, so that the equation we solve is (see numerical
details in the Appendix):
\begin{eqnarray}
\partial_t \psi - \fet{V}_p \cdot \nabla \psi  =(i+\gamma)\left[\frac{1}{2}\nabla^2 \psi +\left(1-g_p\mathcal U_p - |\psi|^2\right)\psi\right], 
\label{eq:ComovingdGPE}
\end{eqnarray}
where the impurity is described by the Gaussian potential of intensity $g_p=0.01$ and effective
size $a=1$ (in units of $\xi$), and is situated in the middle of the domain with the coordinates $(x, y)= (128,64)$ (in units of $\xi$). As an initial condition, 
we start with the condensate being at rest and in equilibrium with the impurity. This is done by imaginary time integration of Eq. (\ref{eq:ComovingdGPE}) for $V_p=0$ and $\gamma=0$. Then, at $t=0$, we solve the full Eq. (\ref{eq:ComovingdGPE}), and as a consequence, sound waves are emitted from the neighbourhood of the impurity. Their speed is determined by the
dispersion relation $\omega(\fet k)$ giving the frequency as a
function of the wavenumber and can be obtained by looking for plane-wave solutions to Eq. (\ref{eq:psi0lin}). If $\gamma=0$, $\omega(\fet k)$ is
given by the Bogoliubov dispersion relation~\cite{BEC-review} $\omega(\fet k) =
k\sqrt{1+k^2/4}$. Note
that the smallest velocity, $c=1$, is that of long wavelengths, and
that waves of smaller wavelength travel faster. For $\gamma>0$, the planar waves are dampened out and
the dispersion relation becomes
$\omega(\fet k)=-i\gamma (k^2/2+1)+\sqrt{k^2+k^4/4-\gamma^2}$. The damping rate is determined by $\gamma$ and increases quadratically with the wavenumber. Also, in this case
all waves have a group velocity faster than a minimum one that
for small $\gamma$ is close to $c=1$.

When $V_p<1$ all the waves escape the neighbourhood of the
impurity (see an example in Fig. \ref{fig:density_gamma}(a))
and are dissipated in a boundary buffer region that has large
$\gamma$ (see numerical details in the Appendix and Supplementary Material \cite{SupplMat}). After a
transient the condensate achieves a steady state. 
Fig.
\ref{fig:density_gamma}(b) shows a steady spatial configuration
for $\gamma=0$ and $V_p=0.9$. Figs.
\ref{fig:density_gamma}(d-e) show different profiles of the
condensate density along the $x$ direction across the impurity
position for $V_p$. The condensate density is depleted near the
impurity due to the repulsive interaction, and its general
shape depends on the speed $V_p$ and thermal drag $\gamma$. If
$\gamma=0$ and $V_p\leq 1$ the density of this steady state has
a rear-front symmetry with respect to the particle position
(see specially Fig. \ref{fig:density_gamma}(d)), so that under
integration in Eq. (\ref{eq:fp2}) the net force is zero. The
presence of dissipation ($\gamma>0$) breaks this symmetry even
if $V_p<1$ so that a net drag will appear in agreement with the
calculation of Sect. \ref{sec:drag}. When $V_p>1$, there are
waves that can not escape from the neighbourhood of the
impurity, forming parabolic fringes in front of it and the  Bogoliubov-\u{C}erenkov wake behind it.
(see Fig. \ref{fig:density_gamma}(c) and
\ref{fig:density_gamma}(f) and Supplemental material \cite{SupplMat}).   The opening angle of the \u{C}erenkov cone is determined by the dispersion relation of the waves with long-wavelength and satisfy the relation $\sin(\phi)= 1/V_p$ as shown in Fig. \ref{fig:density_gamma}(c). It is clear that it narrows when the speed increases. The consequence is that there is a net drag induced by these fringes  even when $\gamma=0$, and that it would eventually decrease at very large velocity as the angle of the wake decreases.  
Similar gringes in the condensate density around an obstacle in supersonic flows has
also been observed experimentally
\cite{carusotto2006bogoliubov}. 
Movies showing the transient and long-time density behaviour for
several values of $V_p$ and at $\gamma=0$ are included as
Supplemental Material \cite{SupplMat}. 
The fluid suddenly starts to move towards the negative $x$
direction, and its density approaches a steady state after the
transient. Note that during all the dynamics, the density
deviation with respect to the equilibrium value $\rho=1$ is
very small, justifying the perturbative approach of Sect.
\ref{sec:perturbation}. The time evolution
for $\gamma>0$ is qualitatively similar to the $\gamma=0$
shown in the movies, except that the waves become damped and
that there is a front-rear asymmetry in the steady state.

Our numerical setup is well suited to measure the force
produced by the perturbation of the impurity on the fluid, i.e.
the self-induced drag. Nevertheless, in the absence of the
impurity the unperturbed state is the trivial $\psi=1$, so that
$\delta\psi_0=0$ and the inertial force is identically zero. In
order to test the accuracy of our expressions for the inertial
force without the need of additional simulations under a
different set-up, we still use the computed condensate density
and velocity dynamics, produced by the impurity introduced in
the system at $t=0$, but we evaluate the inertial force exerted
by this flow on another test particle located at a different
position. In fact, there is no need to think on the flow as
being produced by an impurity: it can be produced by a moving
laser beam that can modelled by an external potential $V_{ext}$
and the only impurity present in the system is the test
particle on which the force is evaluated. In the following we
evaluate the inertial and the self-induced drag forces on the
different particles from the general expressions Eqs.
(\ref{eq:fp_unperturb})-(\ref{eq:fp_perturb}) and from the
approximate expressions of Sects. \ref{sec:inertial} and
\ref{sec:drag}.

\subsection{Numerical evaluation of the inertial force}
\label{sec:numerical-inertial}

We consider a test particle traveling at the same speed $V_p$
as the impurity or laser beam producing the flow, but located
at a distance of 10 coherence lengths in front of it, and 20
coherence lengths in the $y$ direction apart from it. This
distance is sufficient to avoid inclusion of $\mathcal U_p$ or
$V_{ext}$ in Eq. (\ref{eq:psi0lin}) for the neighborhood of the
test particle. Condensate and test particle interact via a
coupling constant $g_p'$ sufficiently small so that the full
force on the later, Eq. (\ref{eq:fp2}), is well approximated by
the inertial part Eq. (\ref{eq:fp_unperturb}), being the
perturbation the particle induces on the flow, and thus the
force (\ref{eq:fp_perturb}) completely negligible.

Figure (\ref{fig:inertial_force}) shows, for different values of
$V_p = 0.1, 0.8, 1$ at $\gamma=0$, the $x$ component of the time-dependent force produced
by the transient flow inhomogeneities hitting the test particle in
the form of sound waves. The size of the test particle, taking
several values, is called $a'$ to distinguish it from the size
$a$ of the particle producing the flow perturbation. Blue lines
are computed from the exact Eq. (\ref{eq:fp2}) or equivalently
from Eq. (\ref{eq:fp_unperturb}) to which it reduces for
sufficiently small $g_p'$. Because of the rather explicit
appearance of the interaction potential in this formula, we
label the blue lines in Fig. (\ref{fig:inertial_force}) as
`potential force'. High frequency waves arrive before
low-frequency ones, because its larger sound speed. We also see
how the frequencies become Doppler-shifted for increasing
$V_p$. We have derived in Sect. \ref{sec:inertial} several
approximate expressions for the inertial force. First, Eq.
(\ref{eq:F0full}) is obtained with the sole assumption (besides
$g_p$ sufficiently small) of smallness of the unsteady and/or
inhomogeneous part $\delta\psi_0$ of the wavefunction, which
allows linearization. Eq. (\ref{eq:Fi_th}) assumes in addition
weak inhomogeneities below scales $a$ and $\xi$, and finally
Eqs. (\ref{eq:InertialSimpleDimless}) and
(\ref{eq:InertialSimpleDim}) (equivalent under the previous
linearization approximation) completely neglects such
inhomogeneities (or equivalently, they correspond to
$a,\xi\rightarrow 0$).  We show as black lines in Fig.
(\ref{fig:inertial_force}) the prediction of this last
approximation, similar to the most standard classical
expressions. Since we have computed the wavefunction
$\psi=1+\delta\psi_0$ in the comoving frame from Eq.
(\ref{eq:ComovingdGPE}), we actually use expression
(\ref{eq:Fi_th}) without the Fax\'{e}n Laplacian terms, with
$\delta\fet\omega^{(0)}=\nabla(\delta \psi_0-\delta
\psi_0^*)/(2i) - \fet V_p$, and $\fet{\dot V_p}=0$. Fig.
(\ref{eq:ComovingdGPE}) shows that the full force computed from
Eq. (\ref{eq:fp2}) is well-captured by the approximate
expression of the inertial force for small test-particle size
$a'$. Accuracy progressively deteriorates for increasing $a'$,
and also for increasing $V_p$, but this classical expression
remains a reasonable approximation until $a'\approx 1$. The
accuracy can be improved by considering higher-order Fax\'{e}n
corrections, Eq. (\ref{eq:Fi_th}), or even better, by
considering the integral form in Eq. (\ref{eq:F0full}). We have
explicitly checked that keeping the full Gaussian integration
in Eq. (\ref{eq:F0full}) but approximating the integrand in the
Bessel integral by its value at the particle position gives a
very good approximation to the exact force even for $a'=1$.

\subsection{Numerical evaluation of the drag force}
\label{sec:numerical-drag}
\begin{figure}[ht]
  \centering
\includegraphics[width=0.7\textwidth]{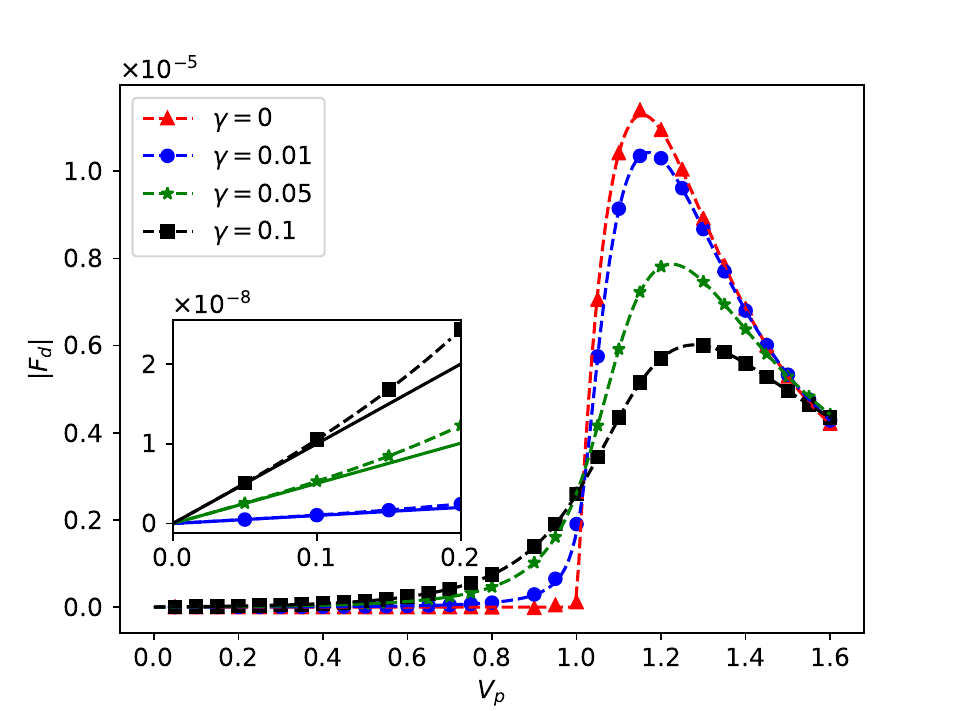}
\caption{Self-induced drag in the steady-state
regime as function of the speed $V_p$. Dashed lines
are the analytical predictions based on Eq. (\ref{eq:forcegamma0})
(for $\gamma=0$) and Eq. (\ref{eq:force_int}) (for $\gamma>0$).
The symbols correspond to the numerically computed force from Eq. (\ref{eq:fp2})
based on direct simulations of the dGPE Eq. (\ref{eq:ComovingdGPE}).
The inset figure shows the small $V_p$ behavior, with solid straight lines giving
the linear dependence of the drag force on the speed
for $\gamma>0$, in the small-$V_p$ approximation given by Eq. (\ref{eq:drag_coeff_full}).
We use $a=1$ and $g_p=0.01$.}
\label{fig:force_drag}
\end{figure}

We now return to the situation in which there is a single
impurity in the system, with size $a=\xi=1$ and $g_p=0.01$. It
moves in the positive $x$ direction with speed $V_p$ producing
a perturbation on the uniform and steady condensate state
$\psi=1$. We compute it in the comoving frame, in which the
particle is at rest and fluid moves with speed $-V_p$, by using
Eq. (\ref{eq:ComovingdGPE}). Since in the absence of the
impurity there is no inhomogeneity nor time dependence,
$\delta\psi_0=0$ and the exact force on the impurity, Eq.
(\ref{eq:fp2}), is also given by the self-induced drag
expression given by Eq. (\ref{eq:density_perturb}). After a
transient, that in analogy with the results for compressible
classical fluids \cite{Longhorn1952,parmar2012equation} we
expect to be of the order of the time needed by the sound waves
to cross a region of size $a$ or $\xi$, the condensate density near
the particle achieves a steady state, and
we then measure the steady drag on the particle. Fig.
(\ref{fig:force_drag}) shows this force, for several values of
$V_p$ and $\gamma$, as dots. 

The approximate value of the drag
force that is obtained under the assumption of small
perturbation (small $g_p$) that allows linearisation is shown
as dashed lines. It is computed from Eq. (\ref{eq:forcegamma0})
for $\gamma=0$ and Eq. (\ref{eq:force_int}) for $\gamma>0$. The
agreement is excellent. As shown in the inset figure, in the
regime of small velocities, the self-induced drag is indeed
linearly dependent on the speed with an effective drag
coefficient that is well captured by Eq. (\ref{eq:drag_coeff_full}). This Stokes-like drag at small
speeds is due to energy dissipation through collisions between the condensate atoms and thermal atoms, quantified by the thermal drag $\gamma$. We notice that the
dependence of the drag force on $V_p$ is consistent with having
a critical velocity for superfluidity even at $\gamma>0$, in
the sense that there is still a relatively abrupt change in the
force (sharper for smaller $\gamma$) around a particular
impurity speed. The superfluidity of BECs at finite temperature
is still an open question. Recent experiments
\cite{sing2016probingsuperfluidity,Weimer2015_criticavelocity}
report superfluid below a critical velocity which is related to
the onset of fringes \cite{wouters2010}. In the dGPE, the
steady state drag is always nonzero. Nonetheless, there is a
critical velocity associated to the breakdown of
superfluidity due to energy dissipation through acoustic excitations. This is the regime where the drag force is
dominated by the interaction of the impurity with the
supersonic shock waves to produce the \u{C}erenkov wake as seen in Fig.
\ref{fig:density_gamma}(c) and observed experimentally
\cite{carusotto2006bogoliubov}. The maximum drag force occurs
near the velocity for which the cusp lines forming the wake
still retain an angle close to $\pi$. With increasing speed,
this angle becomes more acute (Fig.
\ref{fig:density_gamma}(f)), and this lowers the density
gradient around the impurity.

\section{Conclusions}

We have studied, from analytic and numerical analysis of the
dGPE, the hydrodynamic forces acting on a small moving impurity
suspended in a 2D BEC at low temperature. In the regime of
small coupling constant $g_p$ and thermal drag $\gamma$, the
force arising from the gradient of the condensate density can
be decomposed onto the inertial force that is produced by the
inhomogeneities and time-dependence of the condensate in the
absence of the particle, and the self-induced force which is
determined by the perturbation produced by the impurity on the
condensate. When the unperturbed flow can be considered
homogeneous on scales below the particle size and the
condensate coherence length, the classical Maxey and Riley
expression \cite{maxey1983equation}, giving the inertial force
in terms of the local or material fluid acceleration, is a good
description of the force. When inhomogeneities become relevant
below these scales, Fax\'{e}n-type corrections arise, similar
to the classical ones in the presence of a finite-size
particle, but here the coherence length plays a role similar to
the particle size. In addition, the condensate thermal drag
enters into these expressions, at difference with the classical
viscous case. We also determined the self-induced force in the
steady-state regime and shown that it is non-zero at any
velocity $V_p$ of the moving impurity if $\gamma>0$. For small
$V_p$, this force is given as a Stokes drag which is linearly
proportional to $V_p$ with a drag coefficient dependent on the
thermal drag $\gamma$. The energy dissipation associated with this drag is due to the loss of condensate atoms into the thermal cloud and is mediated by the thermal drag coefficient. In this sense, the drag on the impurity relates to the way the condensate dissipates energy at low temperature through particle exchanges with the thermal cloud. We have not considered the additional drag arising from direct interactions of impurity with the thermal cloud, since these are negligible in the low-temperature regime but maybe important at higher temperature. With increasing velocities, there are
corrections to the linear drag and above a critical speed $V_c=1$, the self-induced drag is dominated by
the interactions of the impurity with the emitted shock waves.

We have checked our analytical expressions with numerical
simulations in the situation in which the impurity moves at
constant velocity, possibly driven by external forces different
from the hydrodynamic ones analysed here. When the coupling
constant $g_p$ is sufficiently small so that only the inertial
force is relevant, the equation of motion of the impurity under
the sole action of the inertial force would be $m_p d\fet
V_p(t)/dt=\fet F^{(0)}(t)$, with $m_p$ the mass of the particle
and $\fet F^{(0)}(t)$ one of the suitable approximations to the
inertial force given in Sect. \ref{sec:inertial}. For larger
$g_p$, when the condensate becomes distorted by the impurity,
we have computed the self-induced drag only in the steady case.
In analogy with classical compressible flows
\cite{Longhorn1952,parmar2012equation}, we expect
history-dependent forces in this unsteady situation. The
dependence on the thermal drag, however, would be quite
different from that of viscous classical fluids, because of the
lack of viscous boundary layers in the BEC case.

In this study, we have focused on a small impurity that can
only shed acoustic waves. Another interesting extension of this
would be to further investigate the drag and inertial forces
for larger impurity sizes, which can emit vortices, and study
the effect of vortex-impurity interactions on the hydrodynamics
forces.
Following the recent experimental progress on testing the
superfluidity in BEC at finite temperature
\cite{singh2016probing}, it would be interesting to test
experimentally our prediction of the linear drag on the
impurity due to the condensate thermal drag at small velocities
by using measurements of the local heating rate. For probing the inertial force, it would be interesting to experimentally tracking the position of the impurity during non-steady superfluid flow.


\ack We are thankful to Vidar Skogvoll, Kristian Olsen, Zakarias
Laberg Hejlesen and Per Arne Rikvold for stimulating
discussions. This work was partly supported by the Research
Council of Norway through its centers of Excellence funding
scheme, Project No. 262644, and by Spanish MINECO/AEI/FEDER
through the Mar\'{\i}a de Maeztu Program for Units of
Excellence in R\&D (MDM- 2017-0711).

\section*{Appendix: Numerical integration of dGPE}
\begin{figure}[ht]
  \centering
\includegraphics[width=0.7\textwidth]{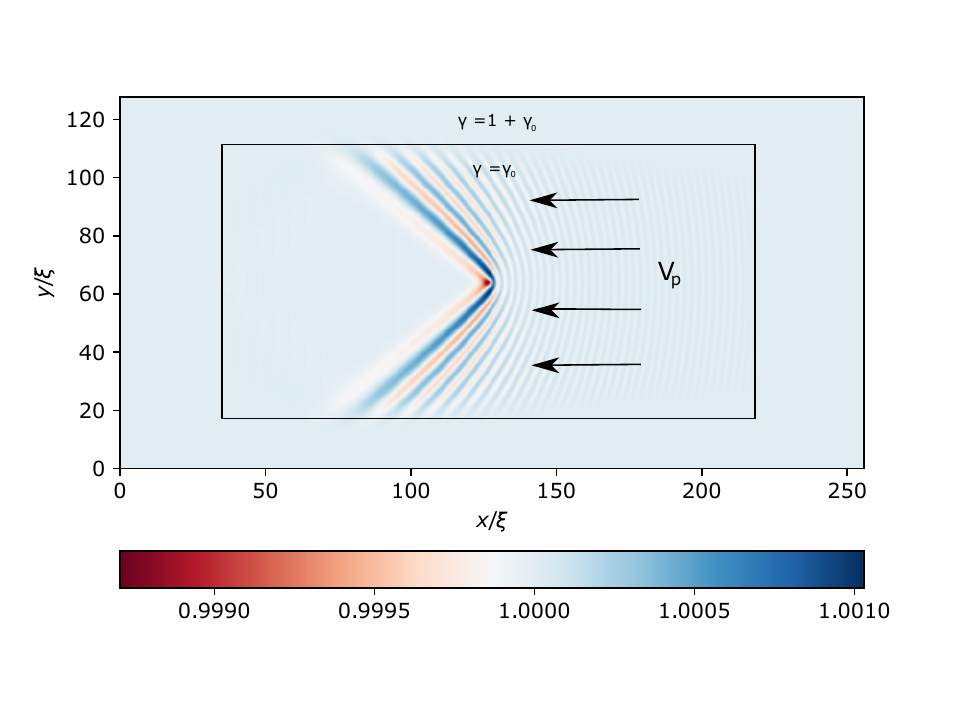}
\caption{Simulation domain showing the buffer region,
outside the main simulation region, in which thermal drag is greatly enhanced
to eliminate the emitted waves sufficiently far from the moving particle (which is at
$x/\xi=128$, $y/\xi=64$). The density shown is the steady state (in
the comoving frame, hence the direction of the arrows indicating the flow velocity in
this frame) for $V_p=1.6$ and $\gamma=0$.
} \label{fig:BufferRegion}
\end{figure}

Numerical simulations of dGPE Eq. (\ref{eq:ComovingdGPE}) are
run for a system size of $128\times 256$ (in units of $\xi$)
corresponding to the grid size $dx=0.25\xi$, and
$dt=0.01\xi/c$. To simulate an infinite domain where the
density variations emitted by the impurity do not recirculate
under periodic boundary conditions, we use the fringe method
from \cite{reeves2015identifying}. This means that we define
buffer (fringe) regions around the outer rim of the
computational domain (see Fig. \ref{fig:BufferRegion}) where
the thermal drag $\gamma$ is much larger than its value inside
the domain, such that any density perturbation far from the
impurity is quickly damped out and a steady inflow is
maintained. The thermal drag becomes thus spatially-dependent
and given by $\gamma(\fet{r}) = \max[\gamma(x),\gamma(y)]$,
where
\begin{eqnarray}
&\gamma(x)= \frac{1}{2}\big(2 + \tanh{[(x-x_p-w_x)/d]}\nonumber\\
&-\tanh{[(x-x_p+w_x)/d]}\big) + \gamma_0,
\end{eqnarray}
and similarly for $\gamma(y)$. Here $\fet r_p = (x_p,y_p)=
(128\xi,64\xi)$ is the position of the impurity and $\gamma_0$
is the constant thermal drag inside the buffer regions (bulk
region). We set the fringe domain as $w_x=100\xi$, $w_y=50\xi$
and $d=7\xi$ as illustrated in Figure \ref{fig:BufferRegion}.

By separating the linear and non-linear terms in Eq.
(\ref{eq:ComovingdGPE}), we can write the dGPE formally
as~\cite{audunsthesis}
\begin{equation}
    \partial_t \psi = \hat\omega(-i\nabla)\psi + N(\fet{r},t),
\end{equation}
where $\hat\omega(-i\nabla) = i[\frac 1 2 \nabla^2+1]+\fet
V_p\cdot \nabla$ is the linear differential operator and
$N(\fet{r},t)=-(i+\gamma)(\mathcal U_p+|\psi|^2)\psi + \gamma
\psi +\frac{1}{2}\gamma\nabla^2\psi$ is the nonlinear function
including the spatially-dependent $\gamma$ and $\mathcal U_p$.
Taking the Fourier transform, we obtain ordinary differential
equations for Fourier coefficients $\psi(\fet{k},t)$ as
\begin{equation}
    \partial_t\hat\psi(\fet{k},t) = \hat\omega(\fet{k})\hat\psi(\fet{k}, t) + \hat N(\fet k ,t),
    \label{eq:DecompositionGPE}
\end{equation}
which can be solved by an operator-splitting and
exponential-time differentiating
method~\cite{cox2002exponential}. It means that we exploit the
fact that the linear part of Eq. (\ref{eq:DecompositionGPE})
can be solved exactly by multiplying with the integrating
factor $e^{-\hat\omega(\fet k)t}$. This leads to
\begin{equation}
    \partial_t \left(\hat\psi(\fet{k},t) e^{-\hat\omega(\fet k)t}\right) = e^{-\hat\omega(\fet k)t}\hat N(\fet k, t).
    \label{eq:ETD}
\end{equation}
The nonlinear term $\hat N(\fet{k},t)$ is linearly approximated
in time for a small time-interval $(t,t+\Delta t)$, i.e
\begin{equation}
    \hat N(\fet k, t+\tau) = N_0 + \frac{N_1}{\Delta t}\tau
\end{equation}
where $N_0 = \hat N(t)$ and $N_1 = \hat N(t+\Delta t) -N_0$.
Inserting this into Eq. (\ref{eq:ETD}) and integrating from $t$
to $t+\Delta t$ we get
\begin{eqnarray}
&\hat{\psi}(\fet k,t+\Delta t) = \hat{\psi}(\fet k,t) e^{\hat{\omega}(\fet k)\Delta t} + \frac{N_0}{\hat\omega(\fet k)}\left(e^{\hat \omega(\fet k) \Delta t} -1\right)\nonumber\\
&+ \frac{N_1}{\hat\omega(\fet k)} \left[\frac{1}{\hat\omega(\fet k) \Delta t}(e^{\hat\omega(\fet k) \Delta t} -1) -1 \right].
\end{eqnarray}
Since computing the value of $N_1$ requires knowledge of the
state at $t+\Delta t$ before we have computed it, we start by
setting it to zero and find a value for the state at $t +\Delta
t$ given that $\hat N(t)$ is constant in the interval. We then
use this state to calculate $N_1$, and add corrections to the
value we got when assuming $N_1=0$.

\section*{References}
\bibliography{ref-2}

\end{document}